\documentclass[fleqn,10pt]{wlscirep}
\usepackage[utf8]{inputenc}
\usepackage[T1]{fontenc}
\title{Controlled kHz laser-driven electron irradiations for pre-clinical applications}

\author[1,+]{C.M.Lazzarini}  
\author[2]{M.Favetta}  
\author[3]{E.R.Szabo}
\author[1]{I.Zymak}  
\author[1]{L.V.N.Goncalves}
\author[1,4]{M.Jech}
\author[1]{S.Lorenz}
\author[1,5]{M.Nevrkla}
\author[1,5]{J.Sisma} 
\author[1,5]{A.Spadova} 
\author[1,5]{F.Vitha}
\author[1]{R.Antipenkov}
\author[1]{P.Bakule}
\author[1]{A.Grenfell}
\author[1]{V.Sobr}
\author[1]{W.Szuba}
\author[3]{J.Dudas}
\author[3]{A.Ebert}
\author[3]{R.Molnar}
\author[3,6]{R.Polanek}
\author[1,7]{S.V.Bulanov}
\author[3,6]{K.Hideghety}
\author[1]{G.M.Grittani}

\affil[1]{ELI Beamlines Facility, The ELI ERIC, Dolni Brezany, Czech Republic}
\affil[2]{U.O.S.D. Fisica Sanitaria e Radioterapia, ASP Trapani, Italy}
\affil[3]{ELI-ALPS Research Institute, ELI-HU Non-Profit Ltd., Szeged, Hungary}
\affil[4]{Faculty of Information Technology, Czech Technical University in Prague, Czech Republic}
\affil[5]{Faculty of Nuclear Sciences and Physical Engineering, Czech Technical University in Prague, Czech Republic}
\affil[6]{Department of Oncotherapy, Faculty of Medicine, University of Szeged, Hungary}
\affil[7]{Kansai Photon Science Institute, National Institutes for Quantum Science and Technology, Umemidai, Kyoto, Japan}
\affil[+]{CarloMaria.Lazzarini@eli-beams.eu}

\keywords{kHz LWFA, electron irradiations, high dose rate}

\begin{abstract}
We report the first in-air irradiations of biological samples with kHz laser-driven electrons with beam energy 20 MeV, high-energy tail extending to 40 MeV, and average dose rate up to 30 Gy/min.
An in-house procedure has been developed to characterize and deliver on-demand (i.e. pre-agreed date and time) the target electron beam energy, dose and dose uniformity.
We present a tolerance analysis on the laser electron parameters, highlighting the importance of beam stability for precise irradiations of \textit{in vivo} zebrafish embryos and \textit{in vitro} U251 glioblastoma cell line. 
The observed improvement in the survival rate of the zebrafish embryos, combined with unchanged cytotoxicity in the cell cultures, indicates promising results for normal tissue sparing while maintaining anticancer efficacy.
The pre-clinical results of this work represent an important milestone towards the clinical translation of laser-plasma accelerators.
\end{abstract}

\begin{document}

\flushbottom
\maketitle
\thispagestyle{empty}

\section*{Introduction}
Cancer could be the leading cause of death by 2040, with 30 million new cases per year \cite{ferlay2024}. Relativistic electron beams are a powerful tool for cancer treatment in radiotherapy, which is nowadays used for half of oncological patients. Medical linear accelerators are already in use \cite{citrin2017}, but limited to 20-25 MeV electrons, due to space limitations imposed by radio-frequency acceleration technology. For this reason, in medical linac, electrons are either used directly for superficial treatments \cite{righi2013} or to generate energetic photon beams for other types of tumors. This process significantly limits the dose rate \cite{lechner2018}. Recently, very high energy electrons (VHEE) from 50 to 250 MeV, emerged as a possible solution to treat deep-seated tumors while keeping the high dose rate. Furthermore, VHEE can have high targeting precision and reduced lateral energy spread \cite{papiez2002,yeboah2002,labate2020}.

At present, the most compact way to generate such electron beams is the Laser WakeField Acceleration (LWFA) technique \cite{tajima1979, esarey2009}. Thanks to the extremely high electric fields sustainable by plasma, it is possible to overcome the fundamental limitation of radio-frequency technology, which is their maximum achievable accelerating gradient ($ \approx 100 \ \mathrm{MV/m} $). The driver of LWFA is usually a high peak power (TW-PW) ultrashort laser (fs), providing VHEE beams from mm-scale targets, with intrinsic pulse duration down to a few fs \cite{lundh2011,horvath2023}. 
LWFA accelerators have been investigated as a potential tool for radiotherapy: \textit{in silico}\cite{fuchs2009,kokurewicz2019}, \textit{in vitro}\cite{mcanespie2025,cavallone2021} and \textit{in vivo}\cite{guo2025}, even though clinically relevant electron beam energy and average dose rate have not been achieved simultaneously.

The advent of optical parametric chirped-pulse amplification (OPCPA) high-power kHz laser systems opens the way to the acceleration of VHEE beams at sufficiently high average current, as we have shown\cite{lazzarini2024} in the ALFA beamline\cite{ALFA}, located at the ELI-Beamlines facility\cite{lazzarini2019,hideghety2025}.
A beam with average current in the order of nA, with an energy > 40 MeV would be beneficial to enable several medical applications as claimed in \cite{brummer2020, svendsen2021}. 

Research on radiotherapeutic effects driven by ionizing radiation at high dose rates (> 40 Gy/s)\cite{matuszak2022} has become more and more important in recent years. While the first experimental evidence of sparing effects from electron beams has been obtained \cite{schuler2017,schuler2022}, the real mechanism behind it remains unknown \cite{taylor2004,jansen2021}. Work is ongoing to understand the role played by the dose rate, for example by the so-called FLASH effect \cite{harrington2019,schuler2022,borghini2024}, and by the peak dose rate \cite{mcanespie2025}, meaning the effect of ultrashort radiation delivery time on the cancer cells. To better explore and validate this effect, a new source of radiation at the fs scale is needed. 
LWFA technology has the advantage of generating such ultrashort electron beam pulses. 

\begin{figure*}[t]
\centering
\includegraphics[width=6.8in]{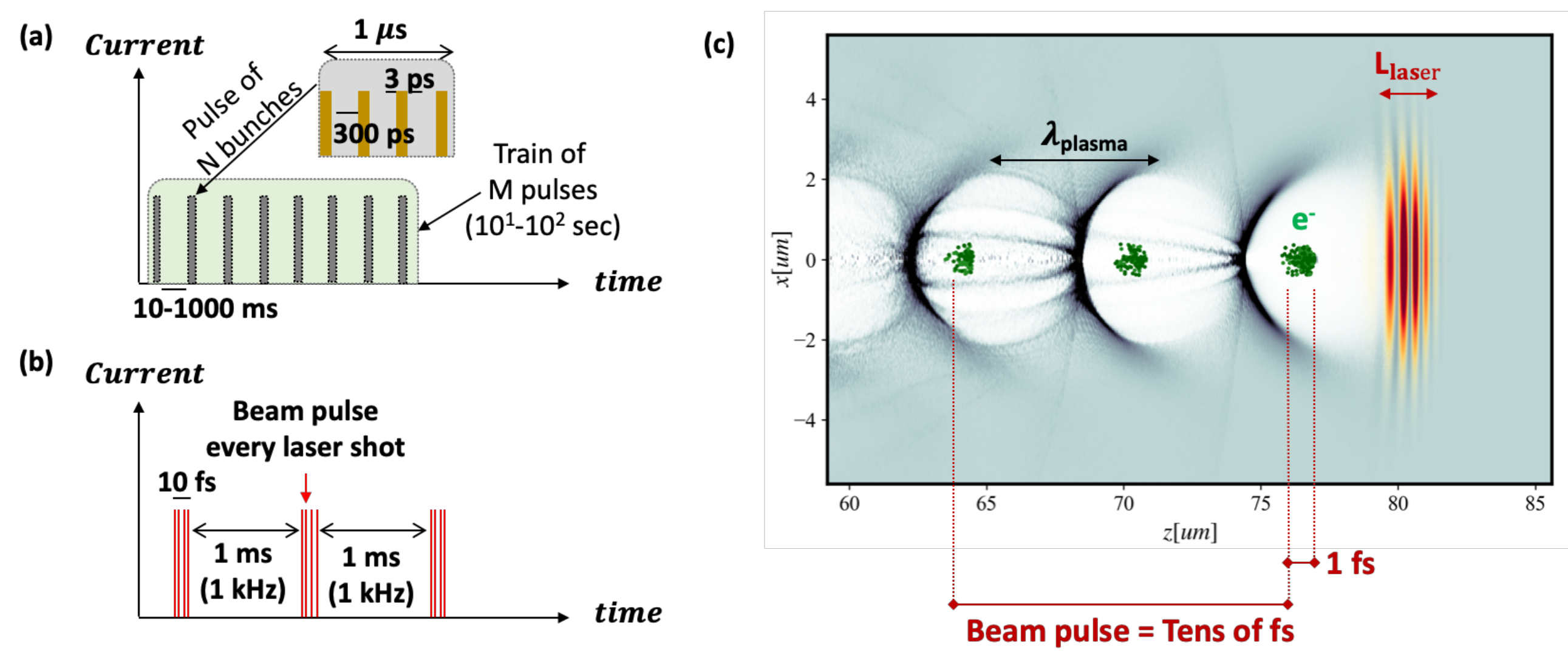}
\caption{Beam time structure. (a) Time structure of electrons accelerated by RF-based technology, where the total dose is given by one or more trains of M pulses, with typical $\mu$s duration and frequency from 1 kHz to 100 Hz, and each pulse is composed of N bunches having individual time duration of a few ps and separated by around 300 ps. In comparison, panel (b) shows the time structure of LWFA-driven electrons where for each laser shot there is an electron beam pulse. (c)
Scheme of the LWFA acceleration where a laser pulse (red) drives plasma waves on its wake (plasma density is shown with light to dark blue) in which are accelerated electron beam bunches (green) with $\mu$m transverse size and fs time duration, forming the beam pulse.}
\label{fig:1}
\end{figure*}

The time structure of high-dose rate beam delivery differs between RF-based and laser-based accelerators.
In the first case, pictured in Fig.~\ref{fig:1}(a), the total deliverable dose from electrons is given with one or multiple trains of M pulses, where pulses typically have a time duration of 1 $\mu$s and a pulse separation form 10 to 1000 ms. Each pulse consists of N beam bunches, where each bunch has a time duration from hundreds of fs to a few ps and the bunch separation on the order of $10^2$ ps. These beams can be delivered with ultra-high doses by a few platforms worldwide, such as CLEAR user facility \cite{lagzda2018}, as well as CLARA\cite{CLARA}, ARES\cite{ARES}, FLASHlab \cite{flashlab}, and Lumitron HyperVIEW\cite{hyperview}. Radiation-induced DNA damage and biological sparing effects have been studied based on these sources, as a function of the tunable dose per pulse and the number of pulses\cite{vozenin2022,wanstall2024}.

Laser-driven acceleration presents a different picture, schematized in Fig.~\ref{fig:1}(b). For each laser shot there is an acceleration event in the plasma target, where each bunch of electrons is injected and accelerated within a period of the plasma wave and can have a time duration as short as a few fs \cite{lundh2011} and a transverse size of a few $\mu$m.
Injection of electrons can happen in different periods of the plasma wave\cite{wenz2019}, as was the case in our previous work\cite{lazzarini2024}, and schematized in Fig.~\ref{fig:1}(c). The total electron beam pulse is then in the order of tens of fs, moving at relativistic speed behind the driving laser. 
In our case, at 1 kHz repetition rate, nothing is coming for the rest of the 1 ms time window. Depending on the acceleration scheme, it is possible to tune the electron beam charge and spectrum, while keeping the same repetition rate.

In this work, we show for the first time, to the best of our knowledge, the delivery of kHz laser-driven electron beams for on-demand in-air irradiation of biological samples (Zebrafish embryos and U251 glioblastoma cell line), with an average dose rate approaching Gy/s.
By on-demand here we mean the ability to start an irradiation experiment at a pre-agreed date and time, with a $\pm 1$ hour accuracy, and a volume uniformity within a 10$\%$ level at the target position. 
For this purpose, we have developed and optimized over three different runs a procedure, described here in detail, with the sequence of the necessary verification steps in preparation for the planned run.

\section*{Methods}
\subsection*{L1 Allegra laser and the ALFA beamline}
The experiment was performed at the ALFA beamline \cite{ALFA} powered by the L1-ALLEGRA multi-stage power-scalable OPCPA laser system as described in detail in Lazzarini et al.\cite{lazzarini2024}. 
The laser upgrade performed in 2025 allows the delivery of  40 mJ in 14 fs (FWHM) at a central wavelength of 830 nm. 
The laser power stability has been demonstrated over different days to be within a few $\%$ over 4 hours of operation; being this the minimum time required to optimize the source and to deliver the planned irradiations. The laser near-field, energy and spectrum before the interaction are continually measured and recorded on a shot-to-shot basis. This ensures the highest electron beam reproducibility, especially for consecutive irradiations.
The laser pulses are focused by a f/5 off-axis parabola (OAP) (specifically dielectric-coated for the laser bandwidth) to a focal spot with an effective waist radius $w_0 = 6 \pm 0.5$ $\mu m$ (Rayleigh length $z_R \simeq 100$ $\mu m$). The focusing was optimized in vacuum and confirmed each day of operation before delivering the electron beams. Considering a focusing efficiency of $> 45\%$ inside the waist diameter, the resulting power on target was $>1.2$ TW  with a peak intensity of $I_p$ = $2 \times 10^{18} W/cm^{2}$. 
The repeatability and stability of the acceleration process can be assured by two factors: first, by the L1-Allegra OPCPA laser intensity contrast lower than $10^{-10}$ at the ps-level; second, by the laser pointing stability on target of $3.6 \pm 0.3$ $\mu$rad (rms-averaged horizontal and vertical). 
We stress the worldwide uniqueness of this LPA system fully composed of reflective dielectric optics, including the chirped mirror compressor. This ensures minimal energy losses and athermal operation, resulting in instantaneous ramp-up and stable delivery of electron beams.

\begin{figure*}[t]
\centering
\includegraphics[width=7in]{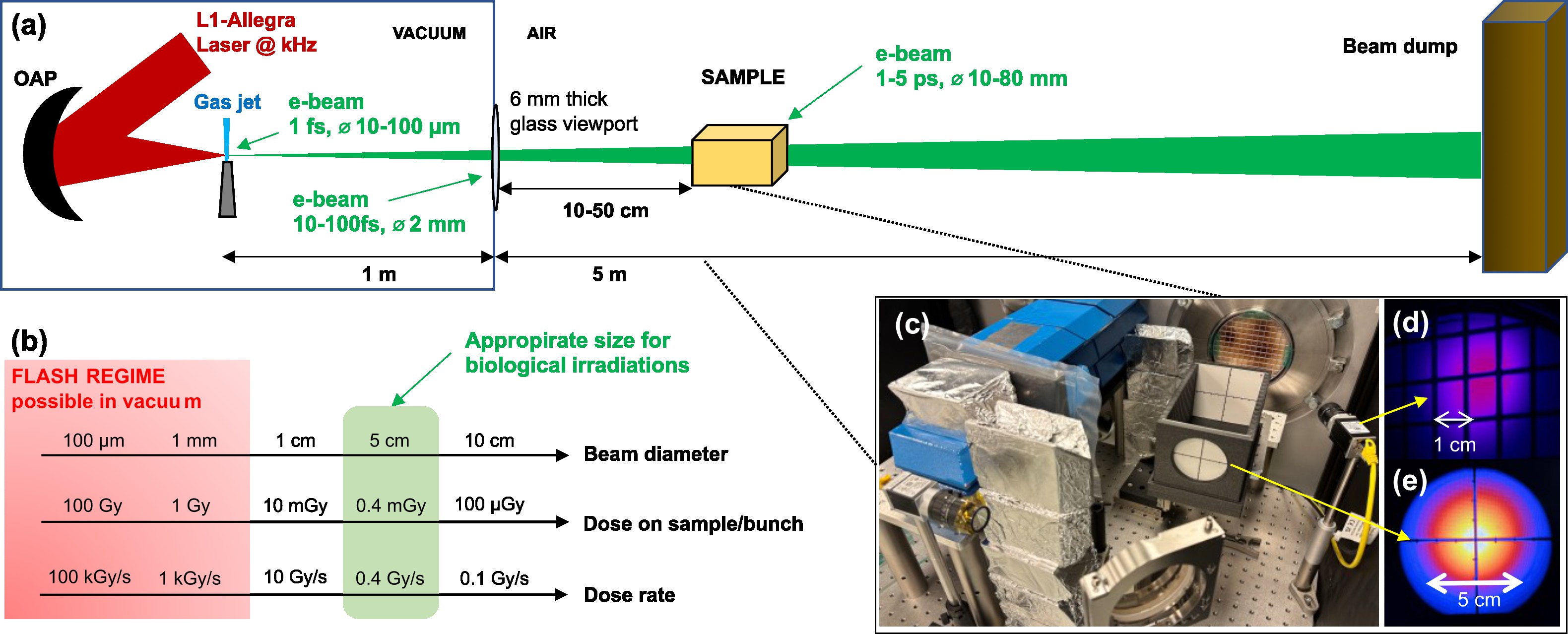}
\caption{Irradiation geometry. (a) Scheme of the accelerator setup for in-air irradiations, with electron beam size and time duration along the path. In yellow is shown the target sample, typically placed 10-50 cm from the vacuum-air interface.  In (b) is shown the relation of electron beam diameters and dose rates. (c) Photo of the in-air table with the target holder and two camera monitoring the Lanex screen attached to the output window. (d) Example of electron beam pointing stability and size observation in real-time during the irradiations with a 1 cm grid and a 2$^{nd}$ calibrated camera. (e) Alignment of the target holder with the main spectrometer camera.}
\label{fig:2}
\end{figure*}

To maximize the electron beam charge at high energy (i.e. > 10 MeV), ALFA has been operated in the high-power mode described in \cite{lazzarini2024}: the gas target was pure nitrogen, produced by a 300 $\mu$m diameter supersonic jet, designed and characterized in-house\cite{lorenz2019,lorenz2020}. 
A fast gas valve synchronized with the laser pulse has been used to control the repetition rate of the electron beam. The continuous flow operation (1 kHz electron beams), required for biological sample irradiation, is achieved with a tailored differential pumping solution \cite{gaspatent} to keep the chamber vacuum level below  $< 1 \times 10^{-3}$ mbar. This is a machine safety requirement, protecting  the laser compressor from contaminations and avoiding over heating of the turbomolecular pumps.

The electron beams are characterized in vacuum using an electron beam spectrometer which consist of a motorized aluminum collimator slit (adjustable from 1 to 5 mm), a motorized permanent magnet dipole (39 mm long, 0.1 T), a LANEX Fast Back scintillator screen, imaged by a 12-bit CMOS global shutter camera. The electron spectrometer has been calibrated in energy and charge with a medical linear electron accelerator\cite{zymak2024}, and also benchmarked with particle tracing from 1 to 100 MeV with both the measured and the simulated dipole magnetic field.

\subsection*{Irradiation Setup}
The electron beams coming out of the gas target travel 1 meter in vacuum, exit the ALFA vacuum chamber through a thin 6 mm glass window and are directed to the irradiation station that is usually aligned with the beam at a distance between 10 cm and 50 cm from the vacuum-air interface, potentially up to 5 m, as schematized in Fig.~\ref{fig:2}(a). 
Typically, all the irradiations with \textit{in vitro} and \textit{in vivo} samples must be performed in air, where it is possible to tune the beam field size ($cm^2$) and uniformity over a fixed area by changing the distance sample-source. This concept is summarized in Fig.~\ref{fig:2}(b), which shows the estimated dose per bunch (Gy) and average dose rate (Gy/s) are estimated as a function of the electron beam diameter. For example, with a typical large area field size of 5 cm, our setup can deliver an average dose rate in the order of  0.5 Gy/s. 
In case ultra-high dose rates are necessary, by having a beam size down to mm range it could be possible to achieve the FLASH regime \cite{vozenin2022}, with an average dose rate > 40 Gy/s and an intrapulse instantaneous dose rate up to TGy/s. In the current setup, this condition might be achieved only in vacuum by placing the sample much closer to the source, to balance the natural spread in time and space, or by introducing additional magnetic focusing elements \cite{whitmore2021,an2024}. 

In Fig.~\ref{fig:2}(c) it is shown our target area for irradiations; where the spectrometer camera is used to carefully align the target holder in 5D by centering the beam on 3 Lanex screens: the main one in vacuum attached to the output window and the other two in air. We were able to align the 15 cm long target support with a precision and repeatability of < 1 mrad (1 mm error), in agreement with the required beam position accuracy for medical applications of $\pm 1$ mm (IEC 60601-2-1). This is of paramount importance for hitting the target while achieving the required dose delivery within  5$\%$ with a dose rate accuracy of $\pm 2 \%$. 

As an additional monitoring tool to observe the beam in real time we installed a second camera off-axis. A grid of 1 cm size is used for a quick visual check of the presence and positioning right before each irradiation (Fig.~\ref{fig:2}(d)). The same diagnostics can be used to address in real time any possible beam drift or loss of charge that might invalidate the current irradiation. This second camera is also used to estimate the dose delivered in real time, based on previous calibrations of the dose rate delivered on target with the charge per pulse retrieved from the main Lanex. This diagnostic can be used to stop the irradiation if the dose delivery differs more than 5$\%$ from the planned value.

To estimate the averaged dose rate, we used target EBT3 Gafchromic films placed on the common support visible in Fig.~\ref{fig:2}(c), and the irradiation time. This ensures an additional verification of the mean beam pointing over the time of the irradiation. The film reading is performed with an Epson Expression 13000XL Pro scanner \cite{ferreira2009}. This allows the reconstruction of the beam dose delivery uniformity and the dose delivered on a target sample, by using the films right in front and at the back of the sample. 

\subsection*{Biological samples}
Wild-type (AB) zebrafish embryos are utilized as a valuable small vertebrate model system for radiobiology investigations, specifically focusing on acute normal tissue reactions. Their small size and high number of offspring are advantageous for the development of novel radiation approaches, enabling meaningful radiobiological studies.
The experiments were performed in accordance with the European Parliament and Council (EU Directive 2010/63/EU). Following irradiation, all biological observations were conducted by the user in Hungary in accordance with the provisions of the XXXIV./72/2022 ethical permit. For each $\textit{in vivo}$ irradiation, about 100 embryos/tube were pipetted into two 1.5 ml Eppendorf tubes containing 1 ml E3 embryo medium \cite{brand2002}.
The biological assessment of survival was performed by observing the heartbeat under a light microscope over the 7 days post-irradiation (dpi). A brief description of the observational protocol, embryo handling, and staging is given in more detail in other works\cite{kimmel1995,brand2002,szabo2018b}.

For the irradiation of the cell culture, the U251 human glioblastoma cell line was purchased from the American Type Culture Collection (ATCC, Manassas, VA, USA). Cells were cultured in Dulbecco's Modified Eagle's Medium (DMEM) High-glucose (4.5 g/L) with sodium pyruvate, supplemented with 10$\%$ fetal bovine serum (FBS), 4 mM L-glutamine and 50 $\mu$g/mL gentamicin. Cells were maintained in a humidified incubator at 37°C with 5$\%$ $CO_2$.
The cells ($10^6$ cells/ml) were then freshly suspended in 1 ml culture medium and were irradiated with the requested of 2 Gy, 5 Gy, 10 Gy dose levels, in 1.5 ml conical Eppendorf centrifuge tubes. The cells in two sample holders were positioned one behind the other and placed vertically in front of the beam. EBT3 foils were placed in front of and behind the samples to measure the delivered dose. 
To assess cell survival following laser-driven electron irradiation, we employed the clonogenic assay, the gold standard method for measuring reproductive cell death after radiation exposure. Specifically, we wanted to examine the U251 cell line's colony-forming efficiency at various dose levels.  The cells were freshly suspended before irradiation and plated immediately after to minimize the disturbance in normal culturing conditions.

\section*{Results and Discussion}
\subsection*{Tolerance analysis for electron beam acceleration}
Most of the irradiation requests at the ALFA beamline are to study radiation-induced effects and survival rate sparing effects on biological samples and radiation to electronics (R2E) effects. 
For both cases, the important parameters are the electron beam current (nA), the achievable dose/rate (Gy/s) and the precision in hitting the target.
In this view, we decided to optimize the laser-plasma interaction by maximizing the laser peak intensity and finding the optimum focusing of the laser inside the gas density profile, by scanning in 3D the gas target.

\begin{figure*}[t]
\centering
\includegraphics[width=6.8in]{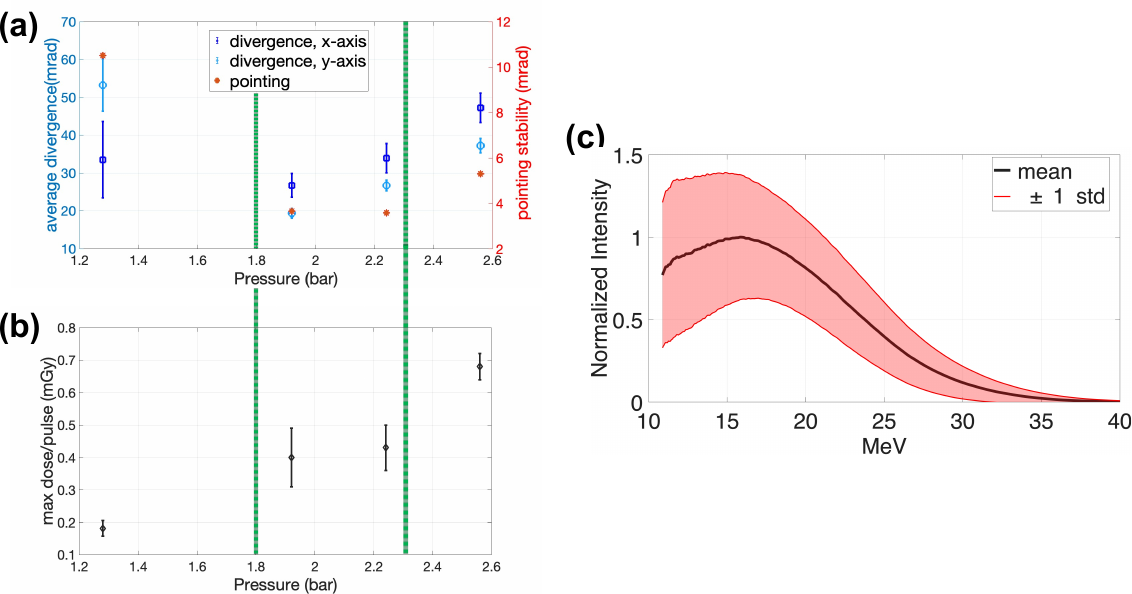}
\caption{(a) Beam divergence for x-axis (square, dark blue) and y-axis (circle, light blue) and average pointing stability (red) as a function of the gas target backing pressure. (b) Measured maximum dose/pulse as a function of the gas target backing pressure. With green lines is shown the optimal range of operation. (c) Example of 1D normalized electron beam spectrum averaged over thousands of pulses, measured at 2.2 bar, corresponding to a plasma density of $n_e = 1.8 \times 10^{19} cm^{-3}$.}
\label{fig:3}
\end{figure*}

\begin{figure*}[t]
\centering
\includegraphics[width=5.5in]{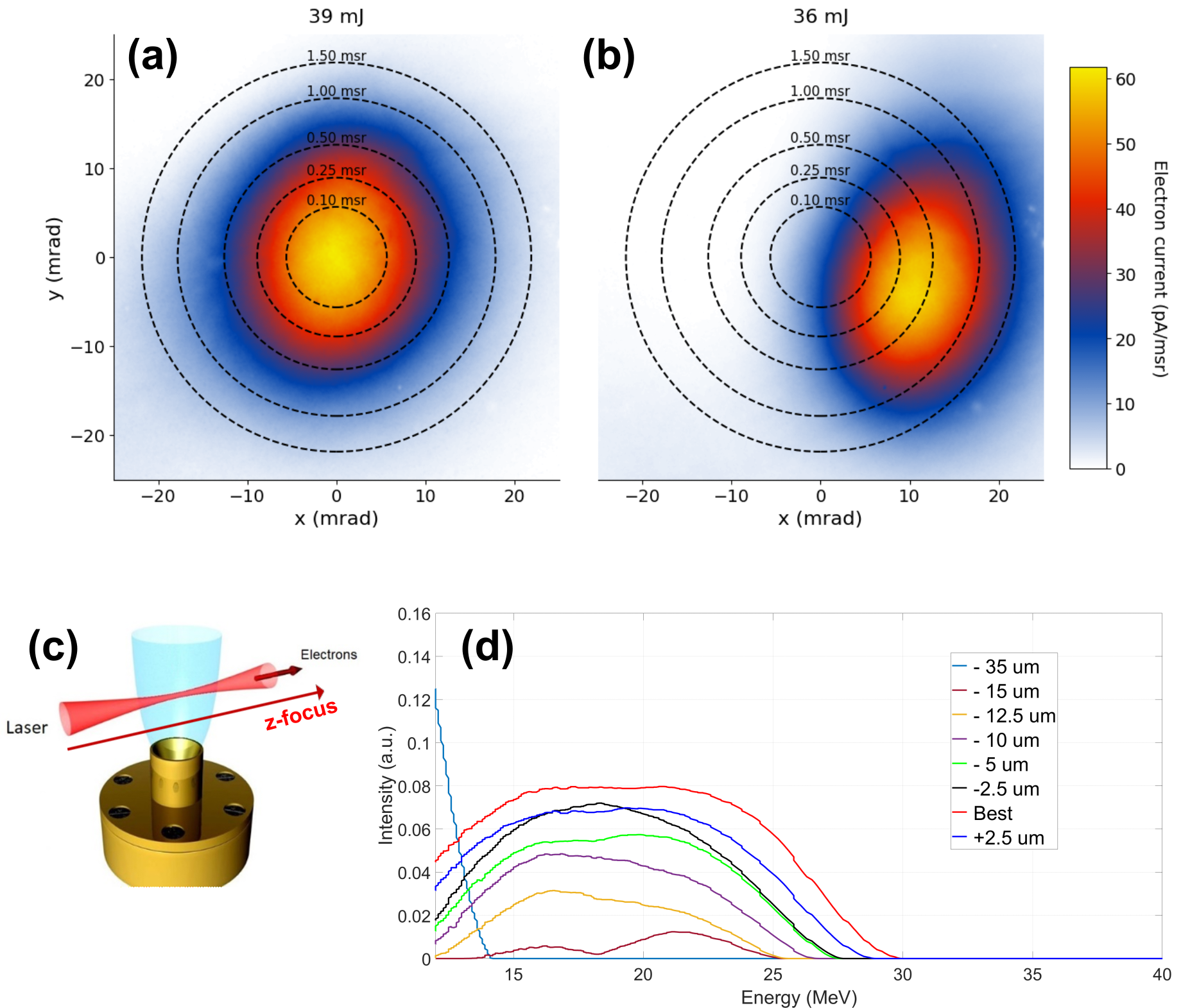}
\caption{Beam stability tolerance on laser parameters. The top panel shows the effect on the beam pointing and beam current for a laser energy drop from 39 mJ (a) to 36 mJ (b), corresponding to $>10\%$ loss in total charge and a shift of 10 mrad. The bottom panel shows the sensitivity on the spectral intensity stability versus the focal spot location inside the gas density profile (c), by scanning the target along the beam direction (z-axis). (d) Optimization of the 1D averaged spectrum in z.}
\label{fig:4}
\end{figure*}

As shown in Fig.~\ref{fig:3}, by reducing the gas target pressure (and thus, the plasma density), the average divergence and the pointing stability of the electron beams improve up to a point where self-injection starts to fail for the given laser power (Fig.~\ref{fig:3}(a)). Furthermore, by reducing the pressure, we observe a reduction of the achievable dose/pulse (Fig.~\ref{fig:3}(b)), caused by fewer injected electrons at lower plasma densities, suggesting the presence of an optimal working range. Within this range is possible to accelerate stable electron beams, with quasi-mono-energetic spectral features, peaked around 20 MeV, as shown in 1D spectrum of Fig.~\ref{fig:3}(c) averaged over thousands of pulses. 

To assure a stable delivery of electron beams, all the laser-plasma interaction must be optimized and kept stable. We identify the 3 main parameters that can lead to failure in delivering the correct dose as: (i) the laser power, (ii) the plasma density and (iii) the laser focus position in the plasma volume.
The first means that the laser power must be kept stable to assure that the electron beam will not change its energy, charge nor its shape. We have verified that an unexpected power drop of the laser output greater than 5$\%$ generates a change in the electron beam profile and pointing; as visualized in Fig.~\ref{fig:4} going from a laser energy of 39 mJ per pulse (a) to 36 mJ (b). 

The plasma density (ii) is kept constant by a pressure regulator with enough throughput to keep the desired gas density at the output of the gas nozzle. The location of the laser focal spot must be optimized within the plasma (iii). This is done in 3D, by moving the motorized target tower along the laser propagation direction, with a  of 2 $\mu m$ precision (depicted in Fig.~\ref{fig:4}(c)), and along the transverse directions, with a precision better than 7 $\mu m$ (i.e., the plasma wavelength at the working pressure). The feature that is being optimized is the output electron energy spectrum. Fig.~\ref{fig:4}(d) shows the dependence of the 1D spectrum on the laser focal spot position in the range $< z_R/2$.

\subsection*{Beam monitoring and dose measurements}
Since the vast majority of biological irradiations, either \textit{in vivo} or \textit{in vitro}, are performed in-air, the beam characterization and the available dose rate assessment must also be verified outside the vacuum after the exit viewport. This is done first by the small Lanex, placed on the supporting structure visible in Fig.~\ref{fig:2}(c), aligned on the electron beam while verifying that the beam pointing and shape do not change over minutes. After that, a couple of gafchromic EBT3 films are irradiated for 10 seconds, separated by 2 cm of PMMA plastic. This allows evaluating the beam uniformity and the achievable average dose rate while considering both, the beam divergence and pointing stability. Typically, we optimize the electron beam to be used for irradiations over an area of $2 \times 2$ cm$^2$, targeting a beam position accuracy of $\pm 1$ mm and a dose uniformity < 2$\%$, to satisfy medical requirement guidelines \cite{IEC}.

\begin{figure*}[t]
\centering
\includegraphics[width=7.2in]{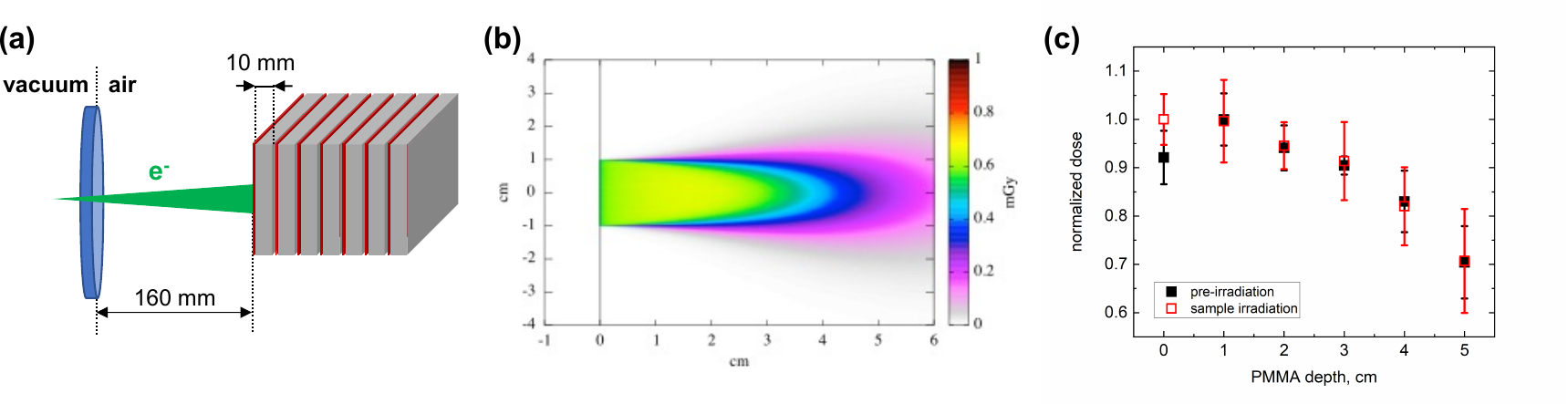}
\caption{Beam dose measurement. (a) Scheme of the dose measurement assembly with a gafchromic EBT3 film in between every 10 mm PMMA slab, taken at a distance > 15 cm from the exit viewport. (b) Monte Carlo visualization of dose deposition for an electron pulse of 1 pC, peak energy 20 MeV and 10 MeV energy spread (as in Fig.~\ref{fig:3}(c)). (c) Averaged percentage depth dose (PDD) curves optimized pre-irradiation (black marker) and validation with the average taken during irradiation days (red marker).}
\label{fig:5}
\end{figure*}

\begin{figure*}[ht]
\centering
\includegraphics[width=7in]{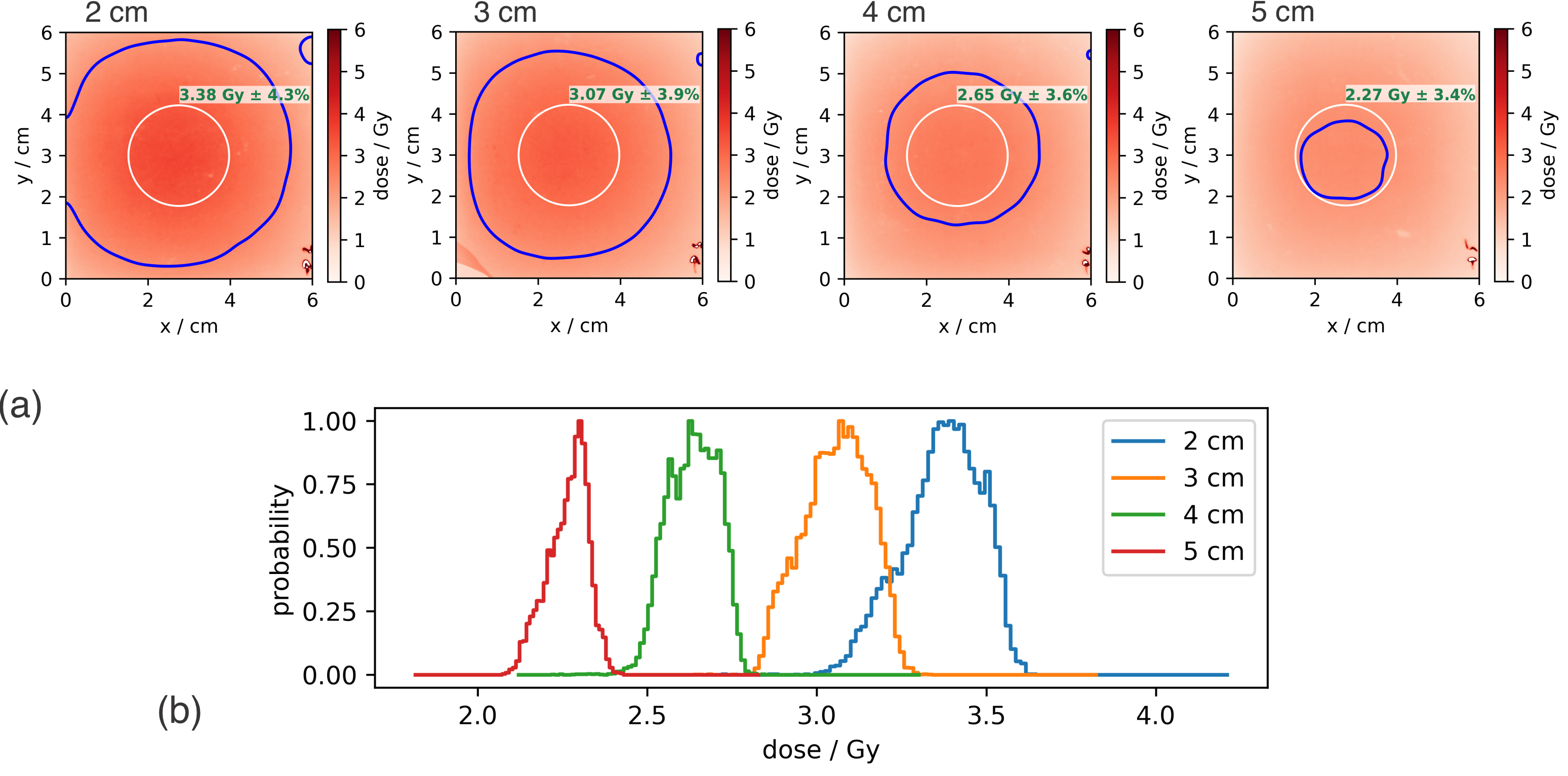}
\caption{Beam lateral dose profile. (a) Scan of films at different increasing depths (in between 10 mm PMMA slabs), where the averaged dose values and standard deviations are given for a 20 mm diameter area (white circle). In blue isodose curves at 2.25 Gy. (e) Corresponding histograms of the dose distribution profile.}
\label{fig:6}
\end{figure*}

Once the target holder alignment is confirmed, the dose deposition at different depths is measured by placing a stack of gafchromic (EBT3 type) films every 1 cm spaced by a water equivalent material, as schematized in Fig.~\ref{fig:5}(a). The dose delivery on a sample depends on the beam energy, energy spread, charge and sample material. In Fig.~\ref{fig:5}(b) we show a visualization of the dose deposition along the propagation direction (longitudinally) and along the transverse direction of the sample area, calculated with FLUKA Monte Carlo, considering a beam of 2 cm size, 1 pC of beam charge per pulse and a spectrum similar to the one plotted in Fig.~\ref{fig:3}(c), i.e. a quasi-monoenergetic distribution peaked at $E_p = 20$ MeV with an energy spread $\Delta E_{FWHM} = 10$ MeV.

Before doing any irradiation, it is important to verify the dose rate, to define the correct dose delivered, at the same distance from the source where the target will be placed. In Fig.~\ref{fig:5}(c), it is shown a typical percentage depth dose (PDD) curve produced by the optimized electron beams pre-irradiation (black curve) and its comparison with the PDD curve taken on the day of the irradiation (red curve). All the curves were measured in water equivalent material under normal atmospheric conditions.

A dose rate of 6 Gy/min can be routinely delivered at 1 cm depth over a circular region of interest (ROI) of 20 mm in diameter. Higher dose rates (up to 30 Gy/min) have been reached over smaller areas and might be available for specific applications.
To ensure that all the samples in the target receive the same dose during each irradiation, it is a fundamental prerequisite to verify the dose delivery uniformity in the 3-dimensional target volume. This must consider the fall-off in all directions. The dose uniformity on target can be tuned by selecting the appropriate distance target to the source. In most of the runs we have obtained a dose uniformity in the irradiation plane $< 5\%$. 

In Fig.~\ref{fig:6}(a) are shown the lateral 2D dose profiles at target depths from 2 cm to 5 cm, where the average dose and standard deviation are calculated from the film scan over the same ROI of 20 mm diameter (white circle).
In Fig.~\ref{fig:6}(b) we show dose distribution histograms inside the ROI, corresponding to the images above for the same irradiation. From this plot, we can understand that a good 2D dose uniformity (here represented by the FWHM of these normalized profiles) can be kept almost constant over the typical depths where the samples can be placed. 

The real delivered dose and dose uniformity are measured for each sample irradiation 24 hours after the irradiation, after scanning all the irradiated films that surround the target.
The advantage of the ALFA beamline consists of the ability to deliver high average dose rate over cm$^2$ area, enabling consecutive irradiations at different doses (e.g., 2, 5, 10 and 20 Gy) within a 30-minute window. Within this timeframe, a set of biological samples can be considered to have been irradiated at essentially the same time.

\subsection*{In-house protocol for on-demand delivery}
\begin{figure*}[t]
\centering
\includegraphics[width=6.5in]{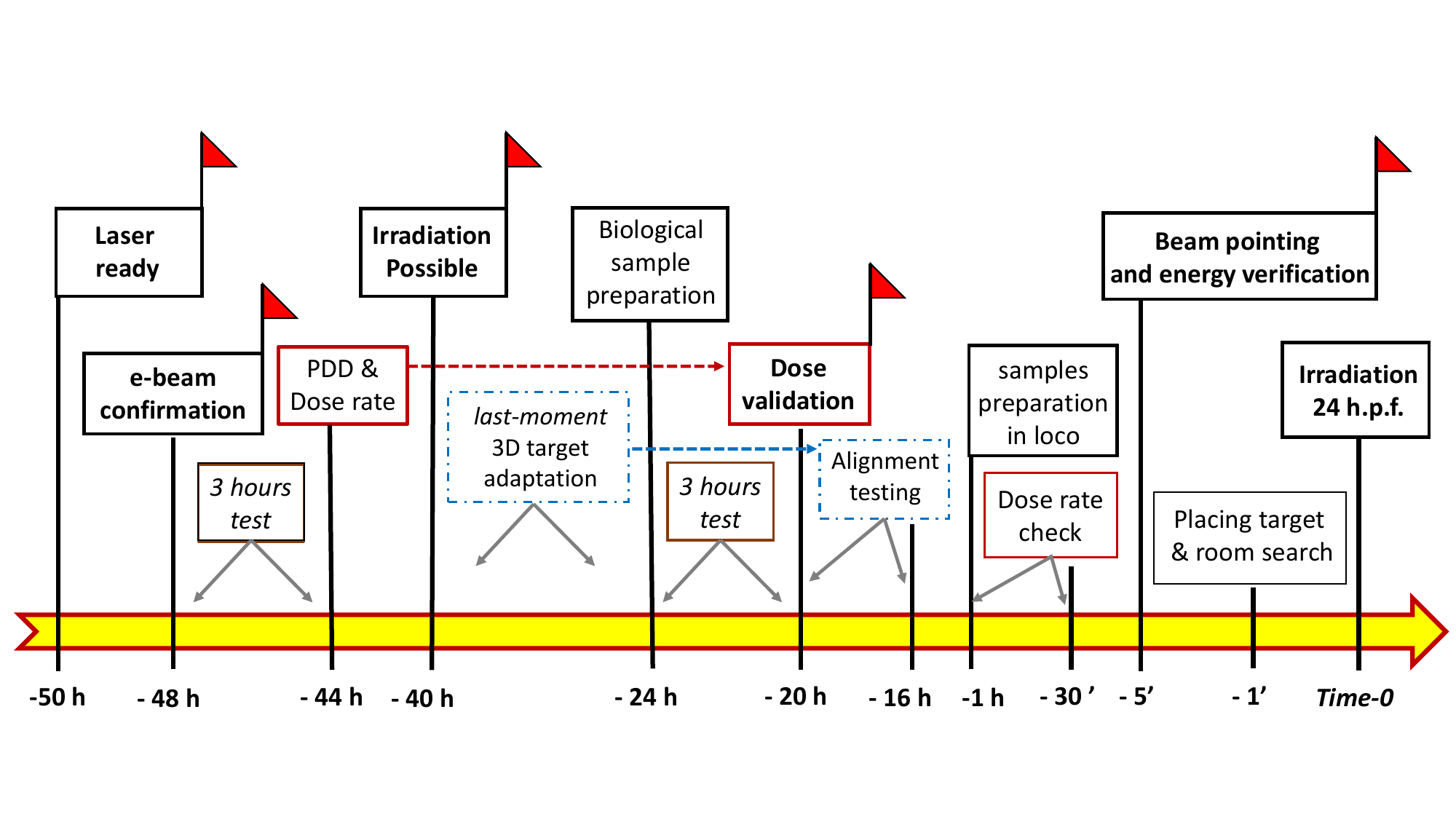}
\caption{Irradiation protocol timeline for the tasks and verifications that are needed to allow on-demand irradiations at a pre-agreed date and time. Red flags indicate the go/no-go checks necessary to confirm the electron beam readiness at target parameters.}
\label{fig:7}
\end{figure*}

The critical challenge to perform what we call here "on-demand irradiations" of biological samples is to deliver specific electron beam parameters at a pre-agreed hour on a specific day, within the actual beam shaping limitations. This request is supported by a particular type of experiments, for example to evaluate the biological response of zebrafish embryos irradiated with electrons as precisely as possible 24 hours post fertilization (hpf). Therefore, logistics plays a crucial role, especially, as in our case, where the biological samples are prepared, transported and managed by an external group coming from a different country.
To satisfy this request at the ALFA beamline we have developed and tested a protocol to be executed in time to ensure the beam is ready when the samples are ready, and not vice versa, being able to confirm the feasibility of the irradiation ideally 48 hours in advance. 
While all the laser-plasma optimization to obtain the desired electron beams should be done days in advance, after fixing the day and hour of irradiation, a series of consecutive verification steps must be followed, as schematized in the timeline of Fig.~\ref{fig:7}. In addition, there are a series of red flags representing key points for the "go" or "no go" of the irradiation.

Let us consider a typical biological irradiation, as the one described below, with the target irradiation to be scheduled on a Wednesday at 10 am, which we call here "Time-0". This imposes that the "biological sample preparation" (here the embryo fertilization) must happen at T = -24 h before that moment and the transfer to the ALFA beamline location should be arranged so that the "samples preparation in loco" happens at T = - 1 h.

The first red flag is the validation of the laser system, to be performed at time T = - 50 h, when all the necessary laser parameters such as near-field energy distribution, far-field for pointing, output energy, pulse compression, and the laser spectrum on target must be verified. After this, within two hours (T = - 48 h), the second red flag consists of the verification of the electron beam parameters: for example, peak energy > 15 MeV, electron beam every shot, beam divergence in the range 10-30 mrad and beam pointing stability < 10 mrad. 
Following, a 3-hour testing of the laser performance can start, in which the laser power stability must be within 2 $\%$, with the electron beam always present.

At the end of this window, a measurement of the PDD curve and dose rate at the preliminary identified sample position is performed. If this step is successful also, meaning that the PDD stack was hit with precision, delivering an uniform dose on the first film, and the electron beam nor the laser system failed during the day, then,  the "irradiation possible" mark can be called at a time T = - 40 h, and the users can be informed that the irradiation will be carried out. 
While a precise evaluation of the PDD can only be inferred after 24 hours, a preliminary inspection of the dose delivered and the dose uniformity is performed to confirm the target design. Indeed, while in the optimal case the user-requested target geometry is defined and agreed weeks in advance based on preliminary measurements, at ELI-Beamlines we have the possibility of 3D printing ad-hoc target holders to be integrated on our 5-axis positioning system. This adaptability has been proven to be possible, adapting the user-requested geometry to the PDD curve estimation up to the time T = - 24 h.

The day before the irradiation the same tests are repeated to verify the laser stability and the day-to-day reproducibility: laser ready at high power at  T = - 26 h, electron beams on specs at  T = - 24 h, followed by a 3-hour full rehearsal of the irradiation. 
The goal of the rehearsal is to improve the beam stability and get the most reliable average dose measurement, upon which the irradiation times will be determined. Based on the previous day's PDD analysis, at T = - 20 h we have the first "dose validation". With this, the irradiation plan is laid out.  After this, no further optimization of the laser and electron beam is possible and the remaining time of the day can be used to refine the target alignment procedure ("alignment testing" in Fig.~\ref{fig:7}). 

\begin{figure*}[t]
\centering
\includegraphics[width=4.5in]{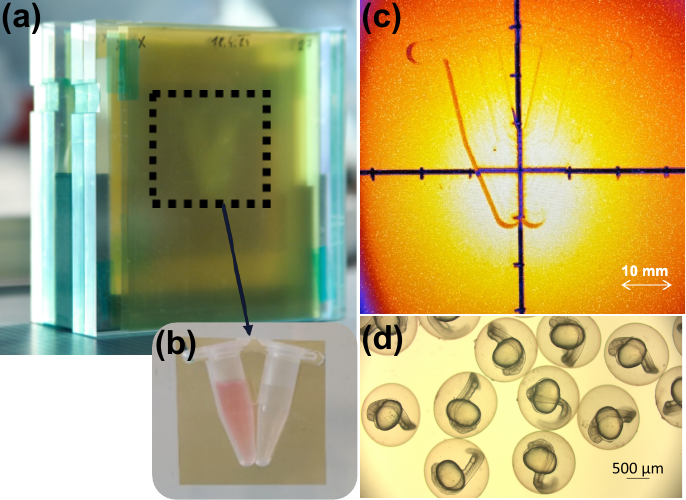}
\caption{Setup for the biological sample irradiations. (a) Target assembly provided by the user with the sample holder positioned within a stack of PMMA slabs (10 x 10 cm) separated by Gafchromic films for dose verification. The holder can accomodate (b) two 1.5-mL Eppendorf tubes containing either cell culture or zebrafish embryos. (c) Example demonstrating sub-mrad precision in delivering the prescribed dose over a 2 cm$^2$ spot size. (d) Representative microscope image of embryos at the time of irradiation.}
\label{fig:8}
\end{figure*}

On the day of the irradiation, the laser must be ready with the expected operational parameters at least by T = -2 h. Similarly, the electron beam by T = -1 h. Between T = - 1 h and T = 30 min, the dose rate is rechecked  by comparing a few gafchromic films with the ones irradiated the day before, for which confirmed measurements exist.
Before placing each target, the beam pointing, size, and energy are verified one last time at T = - 5 min.
If there is any evidence of a change in the electron beam energy, laser energy, or dose rate, the irradiation is postponed and a new PDD evaluation must be performed.
A slight change in the electron beam pointing does not cause a delay, as the target alignment procedure can be executed in only a few minutes.
If everything is ready, T = - 1 min, the target must be placed on the support, with a sub-mm accuracy, and a room clearance sweep is done to permit remote operation.
For each one of the consecutive irradiation, only the last flag (beam pointing and energy verification) is confirmed at T = - 5 min, aiming for the fastest possible manual replacement of the target. Work in ongoing to motorize this step, to further increase its speed and repeatability.

During the irradiation session we are continually monitoring the beam delivery, by observing the presence and pointing of the electron beam on the vacuum lanex (Fig.~\ref{fig:2}d). In addition, we are developing a real-time dose estimation software that will allow us to monitor the delivered dose on target through the measurement of the cross-calibrated charge per shot on the lanex. This tool will allow us to stay within the irradiation dose delivery error of 5$\%$ needed for medical applications. 

\subsection*{\textit{In vitro} and \textit{in vivo} sample irradiation and analysis}
As a first example of the effectiveness of our workflow, we present the first-ever successful \textit{in vivo} (zebrafish embryos) and \textit{in vitro} (human glioblastoma cell lin) irradiation with kHz laser-driven electron beams, conducted as a user experiment at ELI-Beamlines. 
The experiment aimed to examine the dose-dependent survival rates of healthy embryos and the induced morphological abnormalities over 7 days after irradiation. 
Our goal was to deliver precise doses with the highest possible accuracy and uniformity over a volume of $2 \times 2 \times 1$ cm$^3$ at a specific time, ideally 10 am. This requirement was motivated by the need to conduct preclinical investigations into the biological effectiveness of laser-driven electron beams, using embryos irradiated at the early pharyngula stage, ideally at 24 hpf, which allows comparison with the results of previous studies using conventional electron sources. 

In the preparatory phase of the experiment, following a detailed discussion of the experimental conditions, beam parameters, and considering the scientific objectives, a dedicated sample holder was designed (Fig.~\ref{fig:8}(a),(b)) and manufactured by the user to enable precise dose distribution calculations. 
For each \textit{in vivo} run, two tubes with about 100 embryos/tube were irradiated simultaneously by placing them in the specific sample holder inserted into a 10 $\times$ 10 cm stack of PMMA slabs alternated with EBT3 Gafchromic films, as shown in Fig.~\ref{fig:8}(a). This setup allowed us to reconstruct the PDD curves and assess the beam delivery accuracy and uniformity (Fig.~\ref{fig:8}(c)), both on the target surface and along the 10 mm target depth, evaluated by comparing the film directly in front of and behind the sample. We successfully targeted the center with high precision (sub-mrad), delivering most doses within $10\%$ accuracy and achieving less than $10\%$ dose inhomogeneity over the target volume. Several doses were administered within a defined dose range (5-50 Gy) and repeated in subsequent experimental campaigns to obtain relevant conclusions in dose response curves. 

\begin{figure*}[t!]
\centering
\includegraphics[width=4.6in]{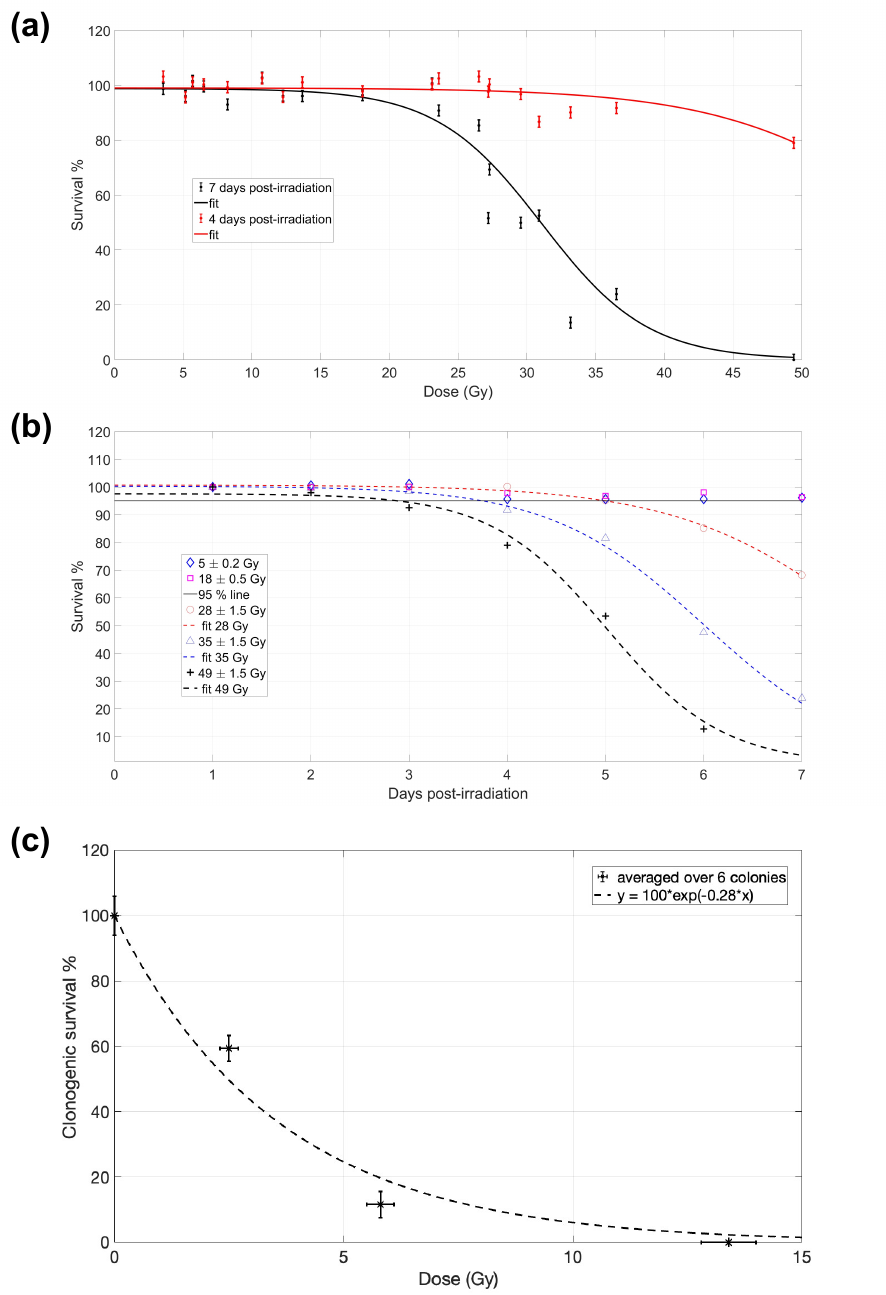}
\caption{Dose- and time-dependent effects of irradiation on Survival. The survival rate of zebrafish embryos (a), (b) was calculated by dividing the number of surviving embryos on each post-irradiation observation day by the total initial number of embryos irradiated. Data points represent the survival percentage normalized to the control survival percentage. (a) Survival percentage at two time points after irradiation. The relationship between absorbed dose (Gy) and overall survival percentage is shown at 4 dpi (red) and 7 dpi (black). The data are fit with: $y_{4dpi}= 99.1/(1+exp(0.14(x-59.3)))$; $y_{7dpi}= 98.8/(1+exp(0.26(x-31.1)))$. (b) Time course of survival decline following various doses. Survival percentage was tracked over 7 days post-irradiation at different dose levels (5-50 Gy). The decay curves for the three highest doses are fitted with: $y_{red}= 1.01/(1+exp(1.04(x-7.7)$, $y_{blue}= 1/(1+exp(1.28(x-6.01)$, $y_{black}= 0.98/(1+exp(1.7(x-5.02)$. In grey the horizontal line for 95$\%$ survival.
(c) Clonogenic survival curve for the U251 cell line, which measures the reproductive capacity of irradiated cells, is plotted against doses (Gy) up to 14 Gy. The data was obtained by averaging six colonies. The curve was fitted with a simple exponential function.}
\label{fig:9}
\end{figure*}

Preliminary results show that using the developed irradiation protocol described previously, we were able to perform three repeatable zebrafish embryo treatments within a time window of a few hours, which yielded relevant biological results for determining the dose-response survival curve shown in Fig.~\ref{fig:9}(a),(b).
To our knowledge, this is the first \textit{in vivo} on-demand delivery ($\pm$1 h) of relativistic-energy laser-driven electron beams at kHz repetition rate, with dose levels and uniformities relevant for future medical applications. 

The survival of zebrafish embryos irradiated with conventional sources remains high up to 10-15 Gy, followed by a sharp, sigmoidal decline, with an $LD_{50}$ of approximately 20 Gy at 7 dpi\cite{szabo2018a,szabo2018b,kari2007}. Meanwhile, doses of 30 Gy and higher proved to be 100$\%$ lethal by day 5 post-irradiation\cite{kari2007,daroczi2006}. Earlier studies using conventional radiation sources and shorter observation periods reported survival rates of 10-75$\%$ after 30 Gy at 4 dpi, whereas 40 Gy or higher was uniformly lethal. 
In contrast, in our LWFA-based irradiation series, survival remained at approximately 100$\%$ in the higher dose range (15-20 Gy). Escalated dose levels (28 Gy, 35 Gy, and 49 Gy) resulted in a time-dependent decrease in survival, but the decline was markedly less steep than what has been described in the literature for conventionally irradiated embryos. At 4 dpi, the embryo survival showed no decline in the LWFA electron irradiated groups at 20 Gy and 30 Gy; whereas survival reported in previously published studies using conventional irradiation sources ranged from 50$\%$ to 75$\%$. The $LD_{50}$ at 7 dpi following LWFA electron-beam irradiation was 30 Gy, indicating an approximately 50$\%$ higher survival compared with published data obtained with conventional sources\cite{szabo2018a,szabo2018b,kari2007}. Supplemented with literary comparisons, these findings suggest an improvement in the survival rate of zebrafish embryos treated with LWFA radiation. 

In addition to the zebrafish model, \textit{in vitro} studies using cultured cells were performed to evaluate the biological effects of laser-driven pulsed electron beams and to validate the effectiveness of our irradiation protocol. U251, a well-characterized human cell line derived from glioblastoma, was exposed to different radiation doses (2 Gy, 5 Gy, 10 Gy) under identical conditions.
Relative clonogenic survival was determined by comparing the survival of irradiated with the control groups, expressed as a percentage in the Fig.~\ref{fig:9}(c). We obtained dose-dependent clonogenic survival curves consistent with published data obtained using conventional sources\cite{howard2017,brunner2021}. This indicates that LWFA irradiation does not exert a differential effect on cancer cells.

\section*{Conclusion}
We demonstrated the possibility of using laser-driven electron beams in the tens of MeV, produced at 1 kHz repetition rate at the ALFA beamline to perform on-demand irradiations targeting specific doses at pre-agreed times. We achieved high precision in hitting the target (sub-mrad), delivering high average dose rates (up to 0.5 Gy/s) with high 2D dose uniformity (< 5$\%$ over cm$^2$) relevant for medical applications such as radiotherapy treatments. We have obtained at best a 3D target dose uniformity and dose delivery error within 10$\%$.
This was achieved thanks to an in-house developed protocol to verify, ahead of the irradiation session, all the requirements on the laser and electron beams in terms of beam energy, average dose rate, dose uniformity and pointing stability.

As a demonstration of the capabilities of our ALFA beamline, we presented the results of the first on-demand irradiations of zebrafish embryos  (24 $\pm$ 1 hpf, \textit{in vivo})  and a human glioblastoma-derived cancer cell line (\textit{in vitro}) using our laser-driven electron beam. Despite the different experimental setups and radiobiological objectives, dose-survival relationships were successfully established in both systems.        The increased survival of healthy vertebrate embryos combined with unchanged cytotoxicity in gold-standard 2D cancer cell cultures suggests a potential normal tissue protection effect while maintaining anticancer efficacy. These encouraging results warrant further investigation into the radiobiological properties and therapeutic potential of LWFA-based irradiation.

\newpage
\section*{Data availability}
The data that support the findings of this study are available from the corresponding author upon reasonable request.

\bibliography{sample}

\section*{Funding}
This work was funded by the National Science Foundation and Czech Science Foundation under NSF-GACR collaborative Grant No. 2206059 from the Czech Science Foundation Grant No. 22-42963L.

\section*{Acknowledgements}
We thank G. Korn for having initiated the kHz research at ELI-Beamlines, leading to the ALFA beamline.
We acknowledge helpful feedback on the manuscript from E. Chacon Golcher.

\section*{Author Contributions}
C.M.L. and G.M.G. had the idea.
C.M.L. ideated the irradiation protocol.
C.M.L. coordinated the whole experiment and beam preparation.
C.M.L., I.Z., G.M.G., M.J. and A.S. run the ALFA LWFA beamline.
R.A., A.G., V.S., A.S., and W.S run the L1-Allegra laser.
E.R.S., J.D. R.P. and K.H. performed the biological experiment. 
M.F. developed the procedures for dosimetric characterization and in-air target alignment.
I.Z., M.J., R.P. and C.M.L. performed the data analysis.
C.M.L. led the writing of the manuscript.
All the authors contributed to the ideas presented in the manuscript or their realization through discussions during the project or review of the manuscript.

\section*{Competing interests}
The authors declare no competing interests.

\end{document}